# Oxidation-assisted graphene heteroepitaxy on copper foil[†]

Nicolas Reckinger,[1,2,*] Xiaohui Tang,[3] Frédéric Joucken[1,2], Luc Lajaunie[4], Raul Arenal,[4,5] Emmanuel Dubois,[6] Benoît Hackens,[7] Luc Henrard,[1,2] and Jean-François Colomer[1,2]

[1]Research Group on Carbon Nanostructures (CARBONNAGe), University of Namur, Rue de Bruxelles 61, 5000 Namur, Belgium.

[2]Department of Physics, Research Center in Physics of Matter and Radiation (PMR), University of Namur, Rue de Bruxelles 61, 5000 Namur, Belgium.

[3]ICTEAM, Université catholique de Louvain (UCL), Place du Levant 3, 1348 Louvain-la-Neuve, Belgium.

[4]Laboratorio de Microscopías Avanzadas (LMA), Instituto de Nanociencia de Aragón (INA), Universidad de Zaragoza, 50018 Zaragoza, Spain.

[5]ARAID Foundation, 50018 Zaragoza, Spain.

[6]UMR CNRS 8520, IEMN/ISEN, Avenue Poincaré, BP 60069, 59652 Villeneuve d'Ascq Cedex, France

[7]NAPS/IMCN, Université catholique de Louvain (UCL), Chemin du Cyclotron 2, 1348 Louvain-la-Neuve, Belgium.

[†]Electronic supplementary information (ESI) available.

[*]E-mail: nicolas.reckinger@unamur.be

We propose an innovative, easy-to-implement approach to synthesize large-area single-crystalline graphene sheets by chemical vapor deposition on copper foil. This method doubly takes advantage of residual oxygen present in the gas phase. First, by slightly oxidizing the copper surface, we induce grain boundary pinning in copper and, in consequence, the freezing of the thermal recrystallization process. Subsequent reduction of copper under hydrogen suddenly unlocks the delayed reconstruction, favoring the growth of centimeter-sized copper (111) grains through the mechanism of abnormal grain growth. Second, the oxidation of the copper surface also drastically reduces the nucleation density of graphene. This oxidation/reduction sequence leads to the synthesis of aligned millimeter-sized monolayer graphene domains in epitaxial registry with copper (111). The as-grown graphene flakes are demonstrated to be both single-crystalline and of high quality.

# Introduction

Chemical vapor deposition (CVD) holds great promises for large-scale production of high-quality graphene. The very low carbon solubility in copper (Cu) makes it a very attractive catalyst for graphene CVD growth.[1] Low-pressure CVD leads to the self-limited growth of a graphene monolayer[2] while the same outcome can be obtained by atmospheric pressure CVD (APCVD) provided that the amount of injected hydrocarbon is carefully controlled (by working with highly-diluted hydrocarbons for instance).[3] However, graphene sheets produced by standard CVD on Cu are polycrystalline[2] and domain boundaries have a detrimental effect on transport properties of charge carriers in graphene.[4] Besides, large-scale single-crystalline monolayer graphene sheets constitute optimal building blocks for artificial layer stacking with a precise control of the interlayer rotation angle. Twisted bilayer graphene, formed from the stacking of two monolayer graphene sheets, has already demonstrated a range of interesting optoelectronic behaviors.[5,6,7,8]

One approach to grow large-size graphene domains is a drastic lowering of the graphene nucleation density by various techniques such as: growth on resolidified Cu,[9] on Cu annealed at high pressure,[10] or by local feeding of carbon precursors.[11] The most popular method benefits from residual oxygen to suppress nucleation by superficially oxidizing Cu.[12,13,14,15,16] A recent study questions the role of Cu oxidation by rather attributing the effect to the removal of carbon contamination on the Cu surface.[17]

The Cu(111) surface plane is very advantageous for graphene synthesis because its hexagonal lattice symmetry matches well the honeycomb lattice of graphene (lattice mismatch of ~4%), thereby enabling epitaxial graphene growth. A few publications (see Table S1 for more details) report on the formation of large Cu(111) grains spanning several millimeters starting from polycrystalline foils,[9,15,16,18] after annealing at temperatures close to the melting point of Cu (1088 °C), where the grain boundary mobility is high. After such thermal treatments, Cu naturally adopts the (111) crystallographic orientation since it is thermodynamically the most stable for face-centered cubic metals.[19,20] By contrast, Robinson *et al.*[20] report the formation of large Cu grains with a dominant (001) texture, the initial main orientation of the Cu foil. Following this line of thought, a novel method has emerged recently, in addition to the "single-domain" one. It consists in orienting the surface of initially polycrystalline Cu foils along the (111) orientation by long (several hours) thermal annealing at more than 1000 °C. The subsequent epitaxial growth of graphene conduces to the alignment of domains which ultimately merge to form a graphene film free of domain boundaries over several dozen centimeter squares.[21,22] Other more expensive, complicated techniques imply directly working on Cu(111) single crystals[23,24,25] or depositing a thin layer of copper epitaxially oriented on α-$Al_2O_3$(0001)[26,27,28] or MgO(111) substrates[29].

The mechanism underlying the growth of large grains in materials is known as abnormal grain growth.[30,31] Cold-rolled metal foils are polycrystalline in nature. After annealing at high temperature, the process of primary recrystallization results in the formation of new dislocation-free grains. When subjected to further annealing at high temperature, the average grain size continues to increase because it is thermodynamically more favorable to reduce the total grain boundary energy. Gradually, larger grains grow at the expense of smaller ones via grain boundary migration (normal grain growth). In specific conditions, normal grain growth may give way to abnormal grain growth where a selective growth of a few "giant" grains occurs by absorbing the small neighboring ones. Abnormal grain growth can only proceed if normal grain growth is somehow inhibited, notably by grain boundary pinning.[30]

In this work, we describe a new method, merging the advantages of the "single-domain" and epitaxial growth approaches, where we take advantage of residual oxygen in the gas phase. First, by slightly oxidizing the Cu surface, we induce grain boundary pinning and, in consequence, freezing of the thermal recrystallization process. Subsequent reduction of Cu under hydrogen suddenly unlocks the delayed reconstruction, favoring abnormal grain growth over normal grain growth. The posterior adjunction of methane leads to the aligned growth of large-area monolayer graphene domains. Compared to the usual hydrogen pre-growth annealing, this oxidation/reduction sequence is thus doubly beneficial: (1) the (111) reconstruction of Cu foils at the centimeter scale is greatly accelerated, allowing graphene domains to grow in an aligned way through epitaxial registry with the Cu substrate and, (2) the oxidation drastically reduces the nucleation density and promotes the growth of millimeter-sized monolayer graphene flakes.

## Experimental

**Graphene growth**

We start from Cu foil pieces with a size of 3×3 cm$^2$ (Alfa Aesar, reference number 13382: 25-μm-thick, purity 99.8%). The Cu pieces are cleaned in a 2:1 mixture of acetic acid and distilled water (see SI section 16 for more details), rinsed in distilled water, and blown-dry with nitrogen. More precisely, the sample is first put on a quartz boat and inserted into a quartz tube at room temperature. An argon flow of 2000 sccm is then fed into the tube for 15 min and the temperature of the hotwall furnace is increased to 1050 °C. Next, the quartz tube is introduced into the furnace and the argon flow decreased to 500 sccm, with the immediate (S#1:H$_2$/S#2:H$_2$) or delayed addition of 20 sccm of hydrogen (S#1:noH$_2$/S#2:H$_2$). After one hour in these conditions, dilute methane (5% in 95% of argon) is injected to grow graphene. One hour later, the quartz tube is extracted from the furnace and left to cool down naturally (fast natural cooling) in the same gas mixture.

**Graphene transfer**

Graphene is transferred onto 300-nm-thick silicon dioxide/silicon substrates or TEM grids (Ted Pella, #01896N) by the widely used method based on poly(methyl methacrylate) (PMMA). After PMMA coating, the Cu foil is first partially etched in ammonium persulfate to remove graphene grown on the backside. After rinsing and rubbing the backside with a cleanroom wiper, the foil's etching is continued in a new persulfate solution. The PMMA/graphene stack is next rinsed thoroughly in distilled water and fished on the wanted support. The sample is left to dry overnight and, finally, PMMA is removed with acetone.

Additional details regarding the experimental techniques (scanning electron microscopy, energy-dispersive X-ray spectrometry, electron-backscattering diffraction, low-energy electron diffraction, X-ray photoelectron spectroscopy, micro-Raman spectroscopy, transmission electron microscopy) can be found in the supplementary information (SI).

**Results and discussion**

The temperature-time diagram shown in Figure 1a summarizes the so-called "standard" (small, misaligned domains) and novel (millimeter-sized, aligned domains) graphene growth conditions, with the corresponding argon and hydrogen flows. The unique difference between the two processes is the pre-growth annealing of the Cu foils: (1) a **reductive** annealing under argon and hydrogen[2,3,4,13] (dubbed S#1:$H_2$/S#2:$H_2$, see Figure 1a for the notations) or (2) an **oxidative/reductive** annealing under argon, then argon and hydrogen[12,13,15,16] (S#1:**no$H_2$**/S#2:$H_2$). The results of the two pre-growth sequences are compared in the low magnification scanning electron microscopy (SEM) images displayed in Figure 1b and c. We observe that the slightly different thermal treatments result in radically dissimilar Cu foil morphologies. It is indeed clearly seen in Figure 1b that the S#1:H$_2$/S#2:H$_2$ Cu foil is polycrystalline, as testified by the heterogeneous SEM contrast rendered by the diverse Cu grain orientations, due to electron channeling. On the other hand, the other Cu foil exhibits a nearly uniform contrast, as in Figure 1c (except for white lines due the rolling striations, darker contrast due to curving of the foil, and a few elongated grains). The systematic use of low magnification SEM to completely sweep the surface of three S#1:noH$_2$/S#2:H$_2$ foils suggests the same qualitative conclusion: the foils present an almost completely unique crystallographic orientation. However, a few residual polycrystalline areas are seen, most of the time on the edges of the foil (Figure S1a and b), or where it is strongly deformed (Figure S1c and d). In line with that observation, the reconstruction proves unstable on the edges of pieces cut from reconstructed 3×3 cm$^2$ samples when they are heated once again in the same conditions (Figure S1e and g). Strain created during the cutting causes the foil to return to a polycrystalline state on the edges. These observations evidence that deformations in the Cu foil play a very important role in the reconstruction. It is also noteworthy that the reconstruction is not purely superficial but rather extends throughout the whole thickness of the Cu foil, as proven by the SEM inspection of both sides of a reconstructed copper foil (Figure S2). In a final control experiment, it is found out that an annealing under argon and hydrogen of at least four hours is necessary to reconstruct a 3×3 cm$^2$ Cu foil in a similar manner to the S#1:noH$_2$/S#2:H$_2$ treatment (Figure S3), meaning that the oxidative/reductive annealing helps to drastically shorten the recrystallization duration and consequently the cost of the whole process.

To unambiguously identify the crystalline structure and determine the size of the Cu grains, we perform electron-backscattering diffraction. First, as illustrated in Figure 2a, the analysis of a 3×3 cm$^2$ S#1:H$_2$/S#2:H$_2$ foil shows that it is mainly (001)-oriented, as already reported by other works for Cu foils subjected to similar treatments [20,32]. Then, a 3×1 cm$^2$

Cu stripe cut from a 3×3 cm$^2$ S#1:noH$_2$/S#2:H$_2$ foil is investigated in three distinct regions (see Figure 2b exhibiting a photograph of the stripe with color-coded circles locating the analyzed areas in the center and on the two edges). The corresponding inverse pole figure (IPF) maps exposed in Figure 1c–h give a confirmation of the large-scale (111) reconstruction of the Cu substrate. More precisely, the uniform color in each out-of-plane IPF map is representative of a completely (111)-oriented Cu surface (Figure 1c–e), while the uniform color of each in-plane IPF map shows that there is no rotational misfit (i.e. no twinning) between the three spots (Figure 1f–h), so the three areas belong to the same crystal. The slight color variations that we can see in Figure 2c,e reflect the uneven topography of the foil (local corrugations or larger-scale creases, similar to Figure 1b), meaning that, in places, the normal to the investigated surface is not at the right angle with respect to the detector. On the other hand, the IPF map in Fig. 2d is very uniform and the crystallographic orientation much closer to the (111) pole because the copper foil in the center is much flatter. The only observed non-uniformities are the elongated grains that could already be seen previously in the SEM pictures in Figure 1c. These small-sized rare grains are identified as (001)-oriented (Figure S4), in agreement with Ref. [21].

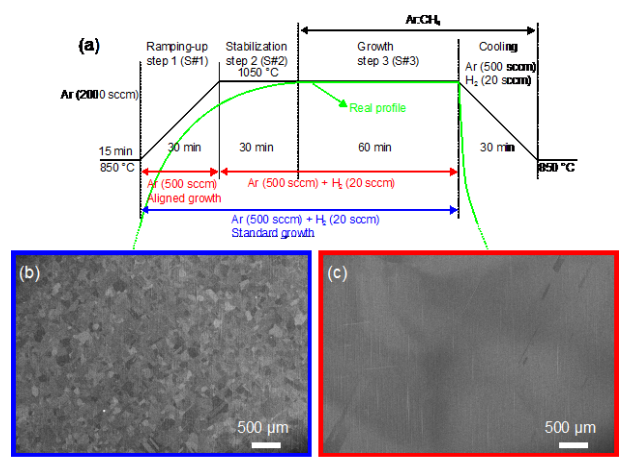

**Figure 1.** (a) Temperature-time diagram summarizing the different steps of the standard (in blue) and aligned (in red) graphene growth conditions with the corresponding argon and hydrogen flows. Scanning electron microscopy pictures of Cu foils after (b) S#1:H$_2$/S#2:H$_2$ (highlighted in blue) or (c) S#1:noH$_2$/S#2:H$_2$ (highlighted in red).

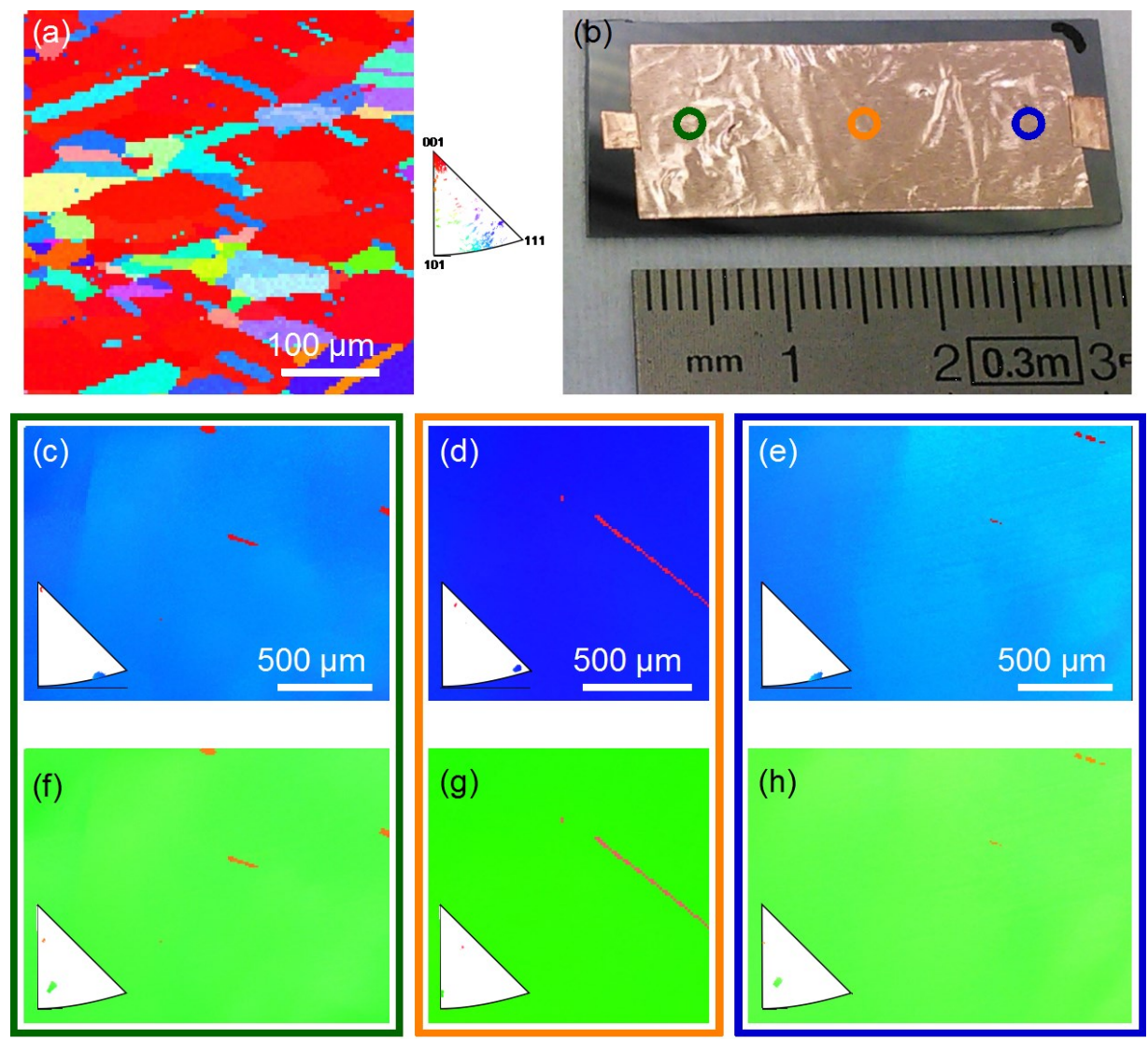

**Figure 2.** (a) Electron-backscattering diffraction out-of-plane inverse pole figure map of a 3×3 cm$^2$ Cu foil after S#1:H$_2$/S#2:H$_2$. (b) Photograph of a 1×3 cm$^2$ Cu stripe cut from a 3×3 cm$^2$ foil after S#1:noH$_2$/S#2:H$_2$, with three colored circles locating where the maps are recorded. Electron-backscattering diffraction out-of-plane (c–e) and in-plane (f–h) inverse pole figure maps at the three locations, with a frame colored according to the circles in (b). In inset: the corresponding stereographic triangles with the 001, 101, and 111 poles.

We now investigate the mechanism responsible for accelerating abnormal grain growth (see an illustration of abnormal grain growth in Figure S5). The sole difference between the two pre-growth annealings considered here is the suppression of hydrogen during S#1. In the absence of hydrogen and its reducing effect, it is known that trace amounts of oxygen are inevitably present in the atmosphere and affect CVD growth.[33,34] In consequence, we are naturally led to believe that oxygen plays a major role in the reconstruction of the Cu foil. To shed more light on the oxidation of Cu, energy-dispersive X-ray spectrometry (EDX) mapping is conducted on a Cu piece after the S#1:noH$_2$ thermal treatment (complementary X-ray photoelectron spectroscopy data can be found in Figure S6a–d). The SEM image in Figure 3a reveals that the surface of the foil is scattered with micrometer-sized faceted inclusions, formed preferentially on the Cu grain boundaries. The corresponding EDX O K and Cu K elemental mappings (Figure 3b and c) disclose that oxygen is almost entirely concentrated in the crystalline inclusions, with a ~33% atomic concentration, matching the stoichiometry of Cu$_2$O; while the dark background corresponds to very weakly oxidized Cu, with less than 1.5% in oxygen (see the corresponding spectra in Figure S7a and b). In Figure 2d, micro-Raman spectroscopy (µRS) further corroborates that the spectrum recorded on the particles corresponds to Cu$_2$O[14,35] and that the background is weakly oxidized. Besides, the Ellingham diagram for the Cu/O$_2$ couple (4Cu + O$_2$ ↔ 2Cu$_2$O) confirms that Cu$_2$O is stable under the considered O$_2$ partial pressure and temperature conditions (see SI section 9 for more details).[36] Finally, in two complementary control experiments, we see that the average Cu grain size is not altered after prolonged (see Figure 3e–g and SI section 10) or repeated (Figure S8a and b) S#1:noH$_2$ annealings, proving that the recrystallization is frozen.

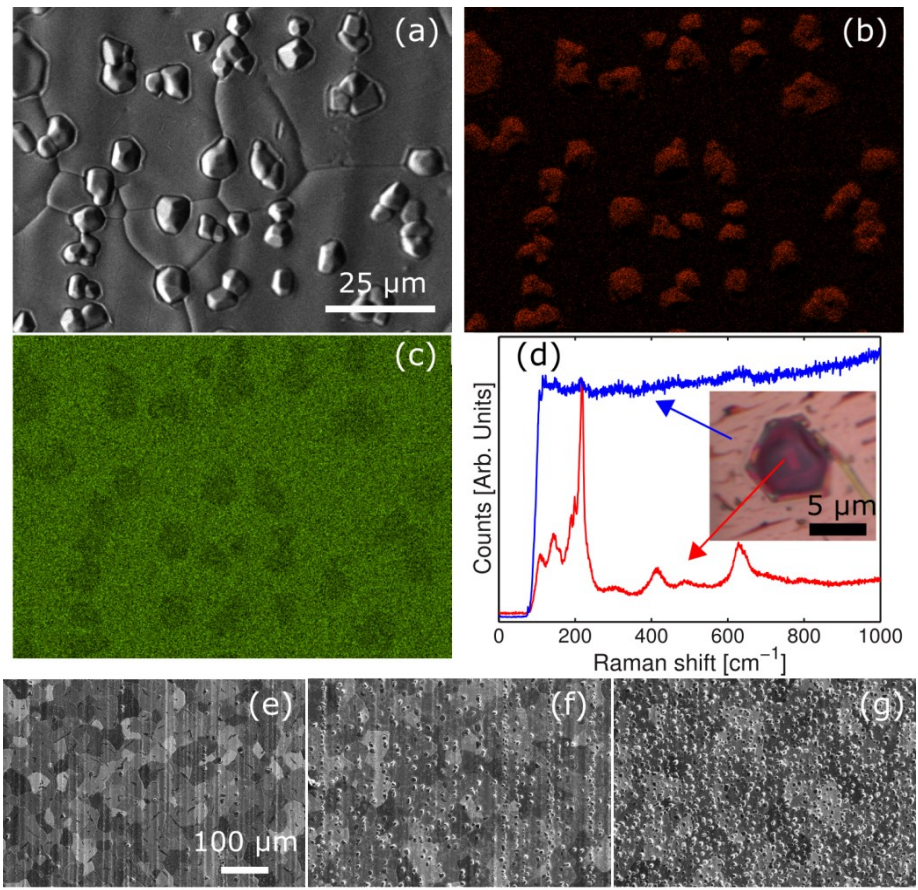

**Figure 3.** Scanning electron microscopy image (a), energy-dispersive X-ray spectrometry O K (b) and Cu K (c) mappings of a Cu foil after S#1:noH$_2$. d) Micro-Raman spectrum of a Cu$_2$O inclusion shown by optical microscopy in the insert. Surface of three Cu foils annealed for (e) 30 min, (f) 1 h or (g) 2 h in argon only.

Based on the knowledge gained through the previous experiments, we propose the following scenario. During the first minutes of annealing under argon, the grain boundary mobility is high enough to enable the average Cu grain size to increase under the effect of the temperature (primary recrystallization). The conditions of temperature and residual oxygen partial pressure evolve in such a way that they become favorable to the formation of Cu$_2$O inclusions. These particles end up pinning down the Cu grain boundaries, even provoking the stagnation of the recrystallization. These conditions are very propitious to abnormal grain growth.[30] So, when hydrogen is introduced in the reactor and the Cu grains are suddenly unpinned, the recrystallization is strongly driven to proceed by abnormal grain growth.

After unveiling the mechanism underlying the trigger of abnormal grain growth, we further investigate the growth of graphene on the (111)-reconstructed Cu foils. The standard conditions (growth following S#1:$H_2$/S#2:$H_2$) lead to misaligned, monolayer graphene domains with a lateral size of 10-15 µm (apex to apex) as can be seen in the SEM image displayed in Figure 4a. This is commonly reported in the literature for graphene synthesized without any special pretreatment of the Cu foil.[3,13] Not content with accelerating the Cu(111) reconstruction, the S#1:no$H_2$/S#2:$H_2$ pre-growth conditions also result in a spectacular size enlargement of the graphene flakes[13] beyond 1 millimeter (compared to 50-100 µm for Brown et al.[21] and Nguyen et al.[22]), and, very interestingly, the different graphene domains are aligned (Figure 4b). By contrast, unlike Zhou et al.[14], it is found that the reverse pre-growth sequence (S#1:$H_2$/S#2:no$H_2$) leads to a much higher nucleation density, resulting in smaller hexagons (< 50 µm) and in multilayer patches (Figure S9). This is most probably due to a much higher number of Cu oxide particles formed during S#2:no$H_2$ which are preferential sites for graphene nucleation.[37] Annealing in hydrogen has indeed a short range polishing effect on the Cu foil,[13,38] smoothing out Cu oxide particles and others defects (rolling striations, etc.), thereby strongly decreasing the nucleation density (Figure S10). The top part of Figure 4c gives an additional illustration of merged graphene flakes aligned growth over ~7 mm. In the inset to Figure 4c, we also show one representative low-energy electron diffraction pattern acquired on the same sample (three others are available in Figure S11). They all display a single set of diffraction spots corresponding to the hexagonal symmetry of both the Cu(111) and graphene lattices aligned with each other. The small lattice mismatch and the relatively large size of the diffraction spots due to the uneven foil prevents from resolving the graphene and Cu(111) diffraction spots, as reported by Brown et al..[21] The diffraction data strongly support the epitaxial alignment between the quasi-monocrystalline Cu(111) foil and graphene.

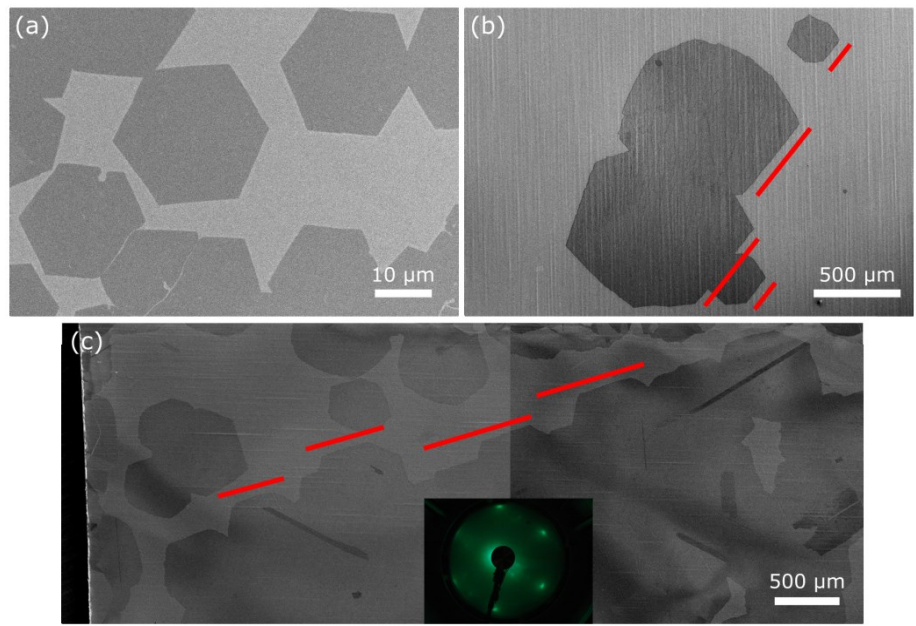

**Figure 4.** Scanning electron microscopy (SEM) image of graphene flakes grown on Cu after (a) S#1:H$_2$/S#2:H$_2$ or (b) S#1:noH$_2$/S#2:H$_2$. c) SEM picture illustrating graphene alignment at the centimeter scale. Inset: low-energy electron diffraction pattern taken on the sample.

Next, we evaluate the structural quality of the as-grown graphene flakes by µRS and transmission electron microscopy (TEM). Figure 5a exhibits a 800-µm-wide hexagonal graphene flake transferred onto a Si/SiO$_2$ (300-nm-thick) substrate. Figure 5b–e present the corresponding µRS mappings of the 2D band full width at half maximum (25.8 ± 1.4 cm$^{-1}$), the 2D-band shift (2684.4 ± 0.8 cm$^{-1}$), the G-band shift (1582 ± 1.1 cm$^{-1}$), and the ratio between the 2D and G integrated intensities (2.3 ± 0.5). The values of each figure of merit match the typical values of high-quality CVD-grown monolayer graphene.[2,13,14] Nanobeam electron diffraction (ED) analyses by TEM are also performed to demonstrate the single-crystalline nature of the graphene flakes.[39,40] Figure 6a shows a 1300-µm-wide hexagonal graphene flake covered by a poly(methyl methacrylate) support film, transferred onto a TEM grid. Up to 25 ED patterns are acquired through the holes of the TEM grid (Figure S12). Figure 6b displays some of the most representative ED patterns. All of them correspond to monolayer single crystalline graphene orientated along the [0001] zone axis. No evidence of extended defects or turbostratic multilayers can be inferred from these data. Figure 6c plots the evolution of the δ angle (angle between the [0-110] plane and the horizontal axis, see Figure 6b) with the covered path. Along the pathway (seen in Figure 6a) covering more than 4.3 mm, the δ values are centered on a mean value of 63.2 ± 0.6º (with a 95% confidence interval). This confirms the good crystallinity and the single crystalline nature of the monolayer graphene flake on its entire area. Both µRS and TEM testify to the excellent quality and spatial uniformity both of the graphene synthesis and the transfer process.

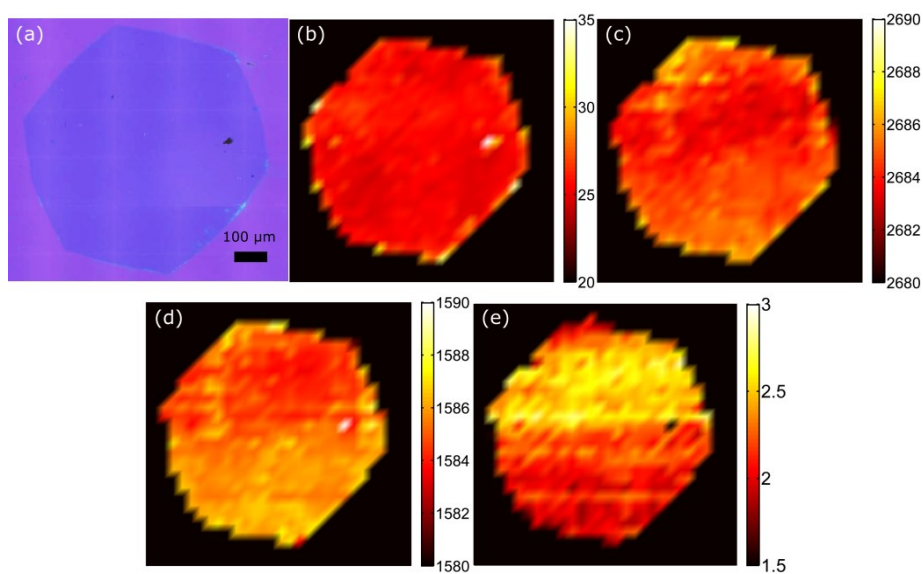

**Figure 5.** a) Optical microscopy picture of a 800-µm-sized monolayer graphene domain transferred onto a Si/SiO$_2$ substrate. Corresponding micro-Raman spectroscopy mappings of the (b) 2D-band full width at half maximum, (c) 2D-band shift, (d) G-band shift, and (e) 2D-band over G-band intensity ratio.

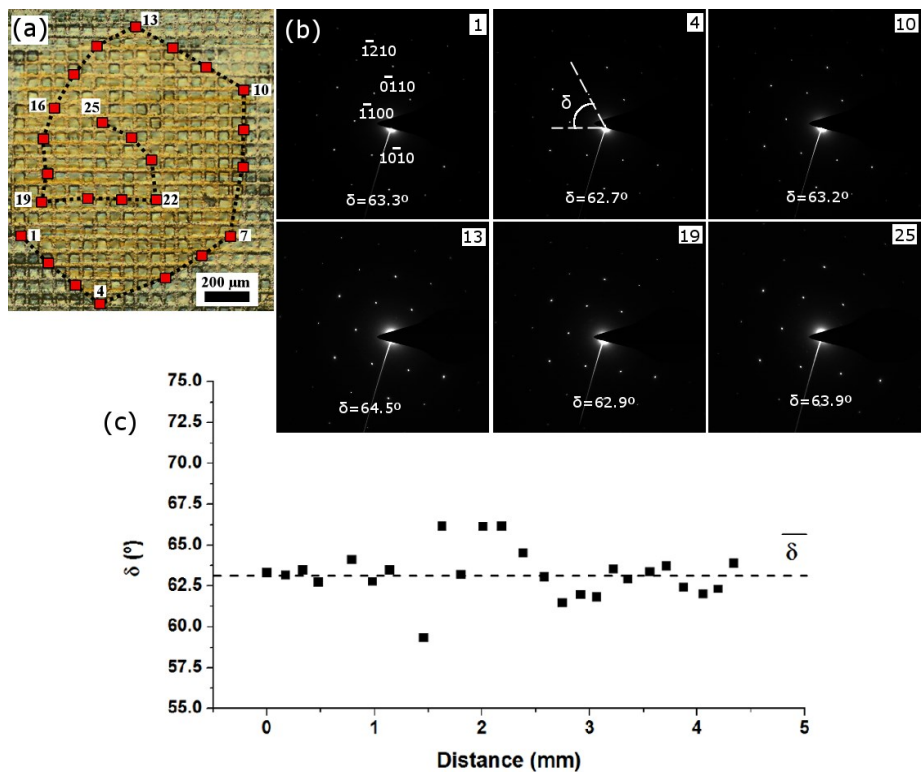

**Figure 6.** a) Optical microscopy image of a 1.3-mm-sized monolayer graphene hexagon covered with poly(methyl methacrylate) transferred onto a transmission electron microscopy grid. Red squares highlight the areas where the 25 corresponding electron diffraction (ED) analyses are performed. b) The most representative ED patterns, labeled according to the analyzed area. c) Variation of the δ angle along the covered pathway shown by the dotted line in (a).

## Conclusions

The present work details an innovative method to grow single-crystalline graphene sheets by CVD, where residual oxygen in the gas phase proves to be doubly advantageous. The superficial oxidation of cold-rolled polycrystalline Cu foils simultaneously (1) accelerates the foil recrystallization by abnormal grain growth in the (111) orientation and (2) drastically lowers the nucleation density. Growing graphene on such Cu(111) templates leads to high-quality monolayer single-crystalline graphene flakes spanning more than one millimeter, aligned at the centimeter scale. These results pave the way to facile, cheap CVD growth of domain-boundary-free graphene films of arbitrarily large dimensions.

## Acknowledgements

The authors acknowledge P. Jacques and A. Vlad for helpful discussions. They also acknowledge P. Louette, F. Ureña, Corry Charlier, and C. Boyaval for their help during the experiments. The research leading to this work received funding from the European Union Seventh Framework Program under grant agreement No 604391 Graphene Flagship. The TEM studies were conducted at the Laboratorio de Microscopias Avanzadas, Instituto de Nanociencia de Aragon, Universidad de Zaragoza, Spain. Some of the research leading to these results has also received funding from the European Union Seventh Framework Programme under Grant Agreement 312483- ESTEEM2 (Integrated Infrastructure Initiative – I3), as well as from EU H2020 ETN project "Enabling Excellence" Grant Agreement 642742 and the Spanish Ministerio de Economia y Competitividad (FIS2013-46159-C3-3-P). This work is partially supported by the Multi-Sensor-Platform for Smart Building Management project (No. 611887).

# Supplementary information

# Oxidation-assisted graphene heteroepitaxy on copper foil

Nicolas Reckinger, Xiaohui Tang, Frédéric Joucken, Luc Lajaunie, Raul Arenal, Emmanuel Dubois, Benoît Hackens, Luc Henrard, and Jean-François Colomer

# Table of contents

# 1) Review on Cu foils with large grain sizes in the context of graphene growth by chemical vapor deposition

| Reference | Orientation | Approximative grain size | Starting foil | Treatment |
| --- | --- | --- | --- | --- |
| [1] | Predominantly (111) | Many millimeters | 99.999%, 0.25 mm thick, Alfa-Aesar | Melted on tungsten; atmospheric pressure, hydrogen and argon flow rates were adjusted to 60 and 940 sccm; temperature was first ramped up to 1000 °C in 50 min and then to 1100 °C in 10 min. The temperature was kept constant for 30 min and then slowly ramped down (1 °C/min) to 1075 °C. |
| [2] | Dominantly (100) | Up to a few millimeters | Alfa Aesar, 99.8% Cu 25 μm thick | Oxidation in air + annealed at 1040 °C in a quartz tube for 3 h in a 10 sccm $H_2$ flow at 150 mTorr. |
| [3] | (111) | Several millimetres | Alfa Aesar, (purity 99.8%, lot no. 13382) | 25 mbar, argon 1000 sccm, 10 min, at 1000 °C. |
| [4] | (111) | A few millimetres | 25 mm in thickness, purity of 99.8% | Thermal annealing at 1000 °C under a $H_2$ atmosphere with 400 sccm flow at 500 Torr. |
| [5] | Predominantly (100) | From a few millimeters to as large as a centimeter | Alfa Aesar, purity of 99.8% | Annealing at 1035 °C in 40 mTorr of $H_2$ for 30 min. |
| [6] | (111) | At the foil scale (up to 16 cm in length, ~2 cm in width) | Nilaco corporation, #CU-113213, 99.9% purity | Annealing it for up to 12 hours at a temperature of 1030 °C in an Ar/$H_2$(100 sccm) environment, a total pressure of 26 Torr. |
| [7] | (111) | At the foil scale (6 cm × 3 cm) | A 100-μm-thick copper foil (from Nilaco, 99.96%) | Annealed at 1075 °C with 1000 sccm Ar and 500 sccm $H_2$ for 2 h, polished using chemical-mechanical polishing/repeated several times until Cu(111) orientation is achieved. |

**Table S1**: Information extracted from several publications dealing with graphene growth by chemical vapor deposition on Cu foils mentioning the formation of large (more than 1 mm) grains, as quoted in the article.

## 2) Impacts of mechanical deformations on the recrystallization

The morphology of three 3×3 cm$^2$ foils is investigated in a systematic way (5×4 mm$^2$ windows every 5 mm, 36 measurement points in total for each foil). We illustrate that mechanical deformations in the Cu foil have a strong influence on the reconstruction. First, after the S#1:noH$_2$/S#2:H$_2$ pregrowth treatment, the Cu foil often remains polycrystalline on the edges, and more specifically around the corners of the Cu foil (see Figure S1a,b), where the foil was stressed during its cutting with scissors from the larger 30×30 cm$^2$ as-received Cu foil. Moreover, it can be seen in Figure S1c that the region around an initial crumple in the foil also remains polycrystalline even after a S#1:noH$_2$/S#2:H$_2$ annealing, while the piece is (111)-oriented everywhere else. Figure S1d shows a closeup view on the crumple with an optical microscope, where the crumple is much better seen than by scanning electron microscopy by the presence of shadows. The same elongated grain is evidenced in both pictures by a red-edged rectangle for facile comparison. These two observations evidence that it would be preferable to receive flat foils, instead of rolled ones, to avoid any manipulation during the unfurling which very often incurs folds and crumples in the foil, notwithstanding pleats already present in the as-received foil. We also believe that the final outcome of the annealing is in a large part determined by the internal strains in Cu foil, which is beyond our control. It is also interesting to note that, due to these "edge effects", the S#1:noH$_2$/S#2:H$_2$ annealing of 1×1 cm$^2$ Cu pieces very often results in a very erratic reconstruction, either absent or very partial.

Finally, we have investigated what happens to a Cu piece cut from a larger (111)-reconstructed one, when it is once again subjected to a S#1:noH$_2$/S#2:H$_2$ pregrowth annealing (see Figure S1e,f). To help the reader to compare, a few identical grains are highlighted in red-edged rectangles. It is revealed that the (111) reconstruction is reversible on the edges of the piece that where stressed due to the cutting. Indeed, in Figure S1g, the top edge that was cut with scissors returns back to a polycrystalline state while the right one (uncut) remains (111)-oriented

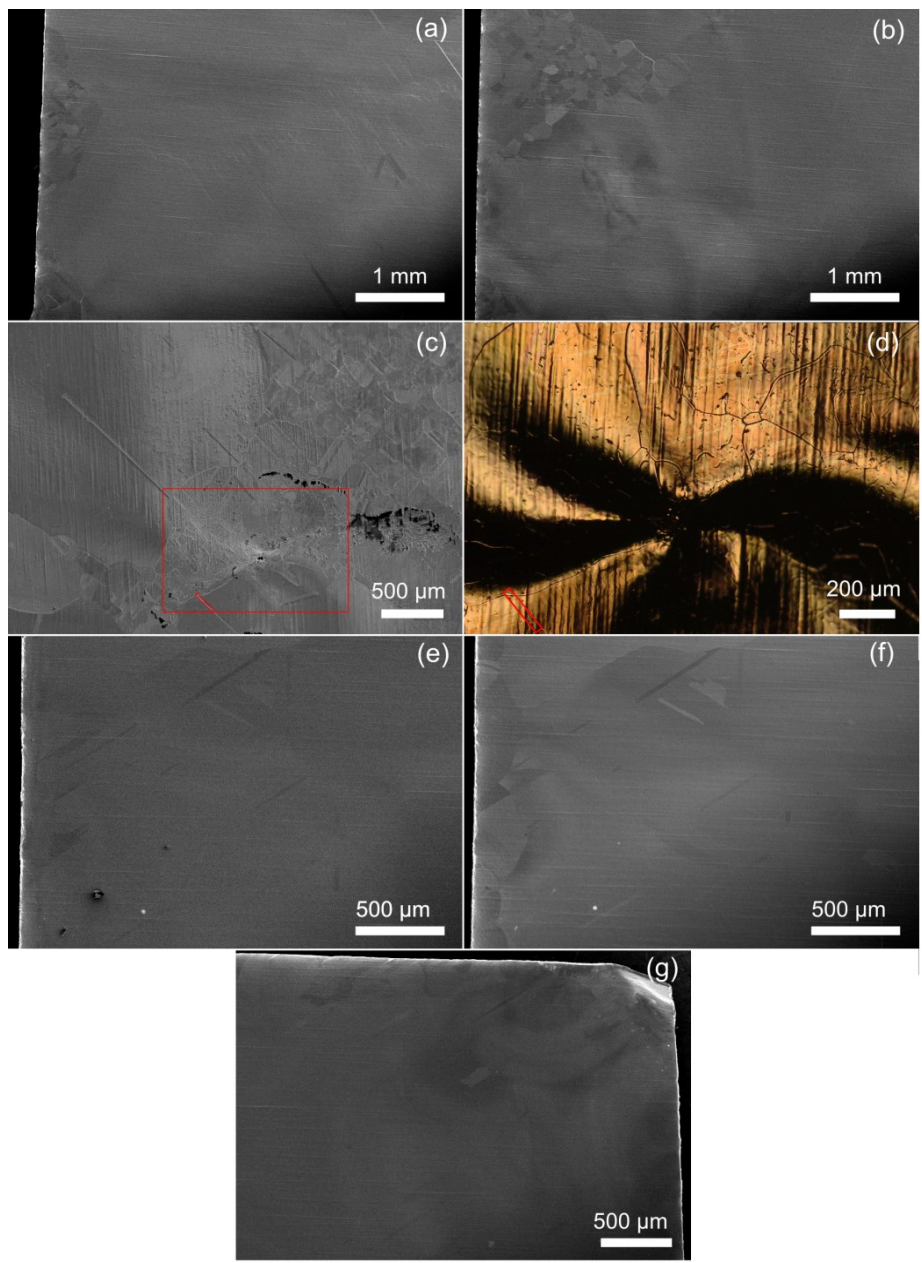

**Figure S1**: Low magnification scanning electron microscopy (SEM) image of (a),(b) the edge of a 3×3 cm$^2$ S#1:noH$_2$/S#2:H$_2$ Cu foil, remaining partially polycrystalline. (c) Low magnification SEM image of a crumpled area in a 3×3 cm$^2$ S#1:noH$_2$/S#2:H$_2$ Cu foil remaining polycrystalline. (d) Optical microscopy picture of the area in the rectangle with red edges in (c), better highlighting the crumple in the Cu foil. (e) Low magnification SEM image of the surface of a ~1 cm$^2$ Cu piece cut with scissors from a 3×3 cm$^2$ S#1:noH$_2$/S#2:H$_2$ Cu foil. (f) Same image taken at the same location after a second S#1:noH$_2$/S#2:H$_2$ treatment. (g) Low magnification SEM image of one corner of the same sample.

## 3) Bulk reconstruction of the Cu foil

The *recto* and the *verso* of the same Cu foil treated by S#1:noH$_2$/S#2:H$_2$ is examined at the exact same location in Figure S2. It clearly appears that both sides are identical even if the contrast of some grains is different due the uneven topography of the Cu foil. To help the reader, a few identical grains are highlighted in red-edged rectangles. The backside picture is flipped vertically to enable a direct comparison.

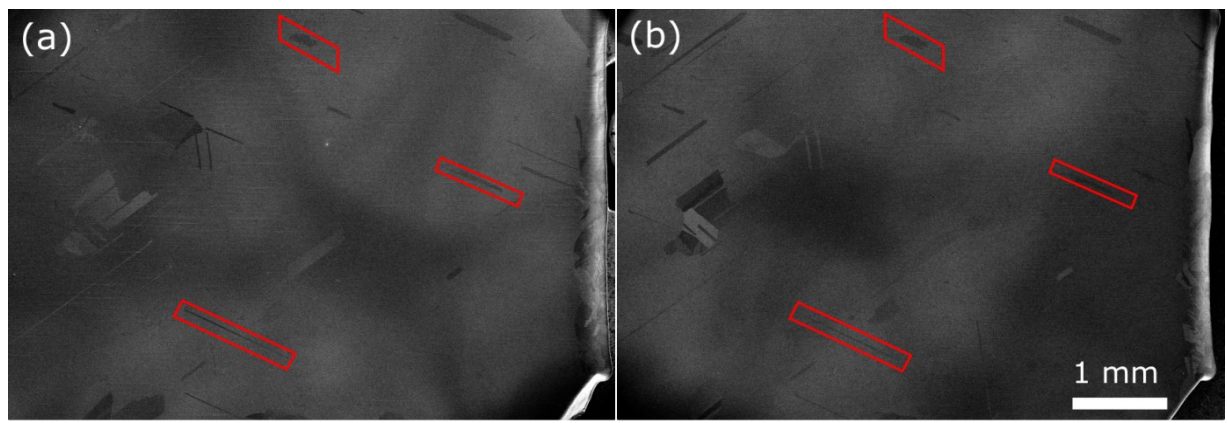

**Figure S2**: Low magnification scanning electron microscopy image of (a) the front- and (b) the backside (flipped vertically) of a Cu foil at the exact same location, evidencing the bulk reconstruction of the foil. The bend corner used to handle the Cu piece is seen to point downward in (b).

## 4) Prolonged Cu foil annealing in argon and hydrogen

Similar to the three 3×3 cm² S#1:noH$_2$/S#2:H$_2$ Cu foils mentioned in the manuscript, this Cu foil annealed in argon and hydrogen during four hours (S#1 30 min + S#2 30 min+ S#3 180 min) is also scanned in 36 positions to get a global view. The two pictures in Figure S3a,b both evidence a uniform contrast (except for the (001) grains and the pleats), testimony to the reconstruction of the foil. Figure S3c reveals that, around a crumple, the foil remains polycrystalline.

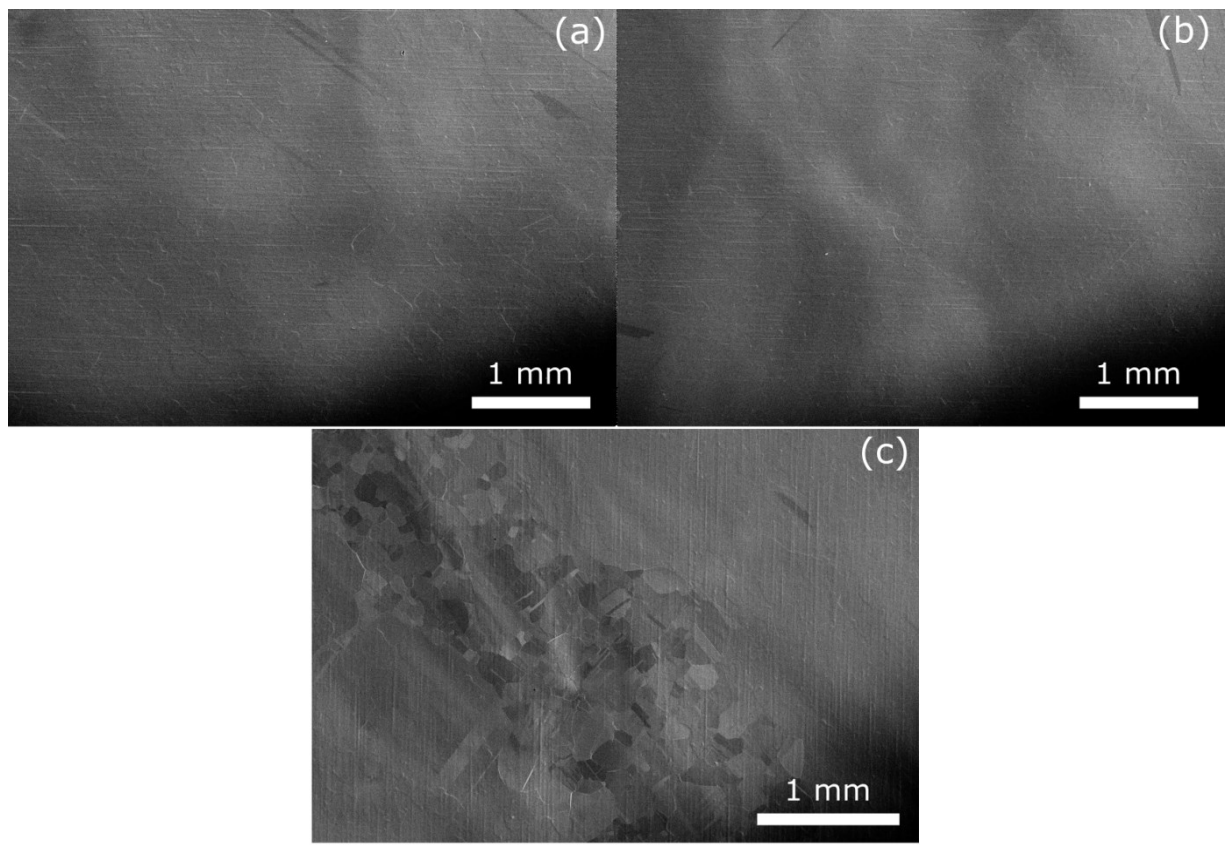

**Figure S3**: Two typical low magnification scanning electron microscopy images of a 3×3 cm² Cu foil annealed during four hours in argon and hydrogen.

## 5) Electron-backscattering diffraction on Cu(001) grains

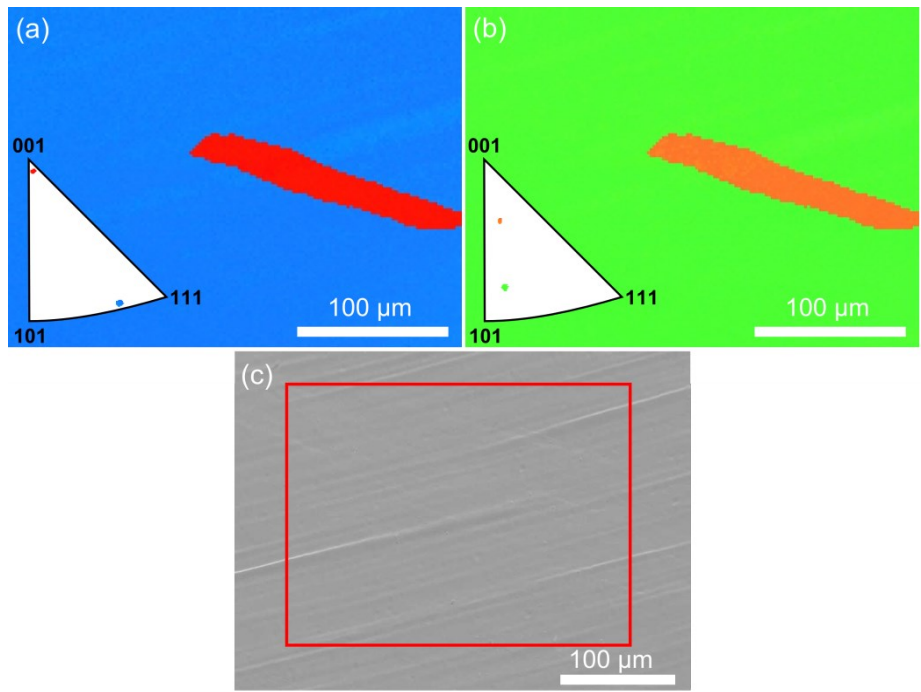

**Figure S4**: Electron-backscattering diffraction inverse pole figure maps of an elongated grain, (a) out-of-plane (*z* direction) and (b) in-plane (*y* direction). In inset are seen the corresponding stereographic triangles with the 001, 101, and 111 poles. The grain has a (001) orientation. (c) Corresponding analyzed area in the red rectangle of the scanning electron microscopy picture.

## 6) Abnormal grain growth

As it is, a cold worked metal foil contains a large amount of dislocations introduced during plastic deformation. As pictured in Figure S5a, when exposed long enough at high temperature, the metal foil becomes fully recrystallized (primary recrystallization), meaning that new dislocation-free grains have formed. Upon further annealing, the average grain size keeps growing in a continuous manner, a process called normal grain growth (NGG, see Figure S5b). In peculiar circumstances (see Figure S5c), NGG can lead to abnormal grain growth (AGG) (also called secondary recrystallization). Contrary to NGG, AGG is a discontinuous phenomenon, resulting in a bimodal grain size distribution. More details can be found in Ref. [8]. The mechanism of AGG is clearly identified in the pictures shown in Figure S5d,e.

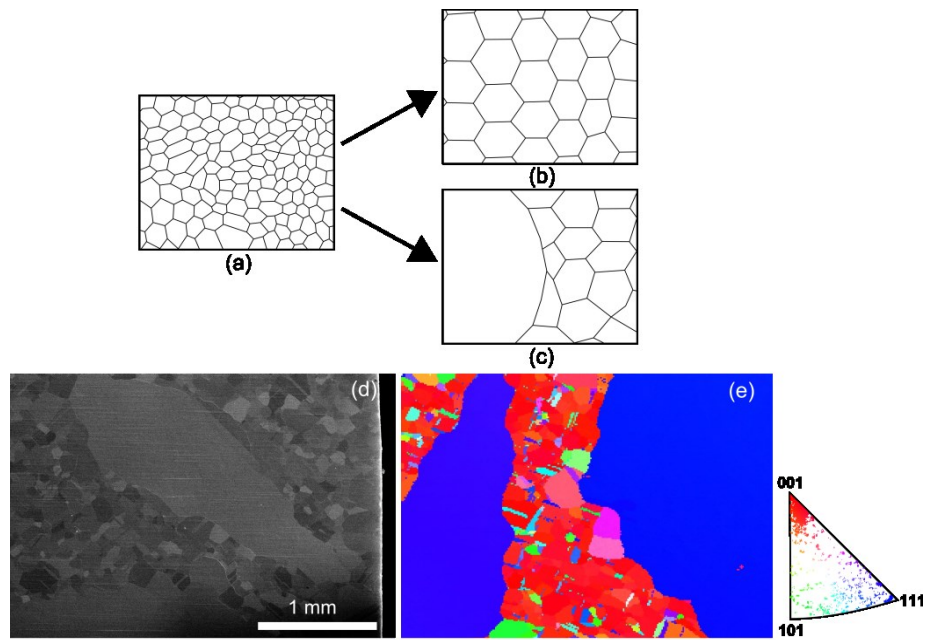

**Figure S5**: Schematic representation of (a) the fully recrystallized state, (b) normal grain growth, and (c) abnormal grain growth. (d) Illustration by scanning electron microscopy of the process of abnormal grain growth, where a large, elongated Cu(111) grain (near the edge of the foil) is surrounded by much smaller grains, and is on the verge of contacting another much larger Cu(111) grain. (e) Electron-backscattering diffraction is used to determine the surface orientation of both grains (same magnification as in Figure S5d). The corresponding stereographic triangle can be seen with the 001, 101, and 111 poles.

S9

# 7) X-ray photoelectron spectroscopy data

Depth profile X-ray photoelectron spectroscopy (XPS) is performed to compare two foils subjected to S#1:$H_2$/S#2:$H_2$ and S#1:no$H_2$/S#2:$H_2$, respectively. The analyses point out that there is no difference between the two samples: at the surface, peaks related to CuO, $Cu_2O$, $Cu(OH)_2$, and $H_2O$ show up with practically the same relative intensities, subsequent to interaction with oxygen and water present in the air[9] (see Figure S6a-d); in depth, Cu is not oxidized at all. It is worth noting that the manufacturer specifies a trace oxygen concentration of 0.01% in the Cu foil, well below the detection threshold of XPS. Next, the same XPS analysis is conducted on a Cu piece after the S#1:no$H_2$ thermal treatment, evidencing that, on the investigated thickness, Cu is weakly oxidized and the atomic oxygen concentration decreases progressively (see Figure S6e). The signal is integrated over a 250-μm-diameter spot and gives thus only average concentrations.

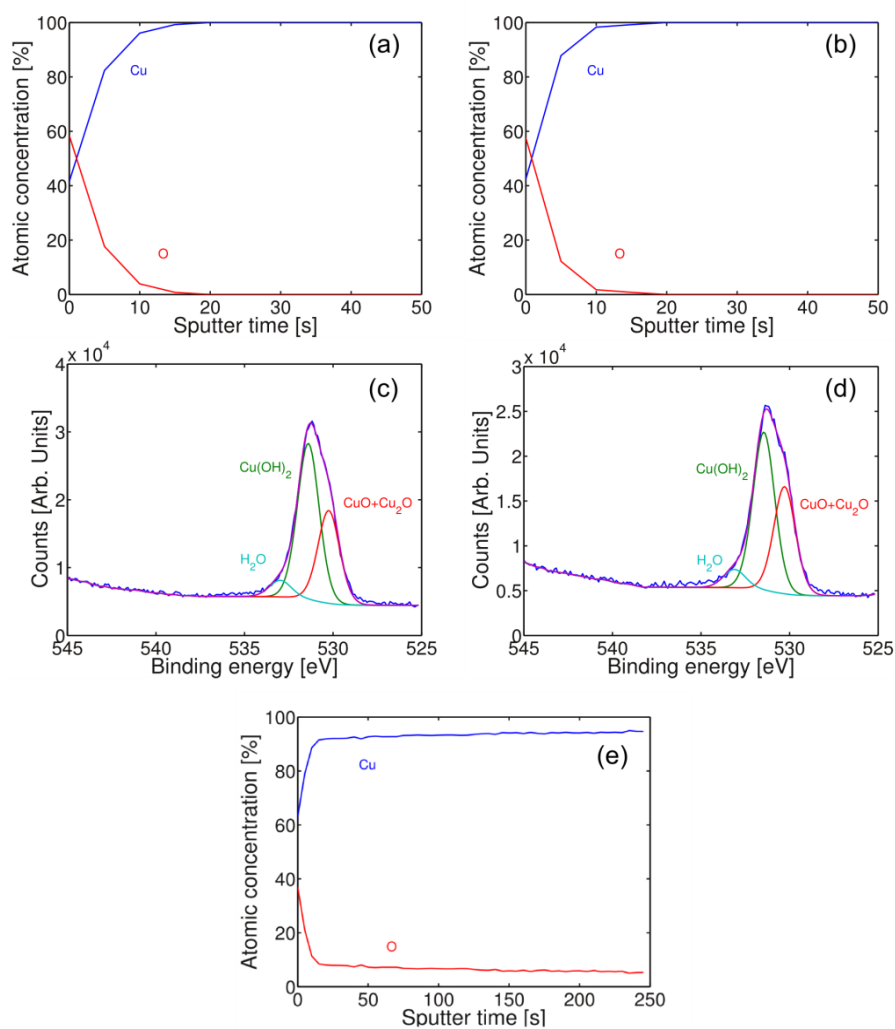

**Figure S6**: X-ray photoelectron spectroscopy (XPS) depth profiles performed to compare two foils subjected to S#1:$H_2$/S#2:$H_2$ (a) and S#1:no$H_2$/S#2:$H_2$ (b), respectively. Cu is oxidized only at the surface of the samples. (c,d) Corresponding high-resolution core level spectra at the very surface of the two foils. XPS depth profiles carried out on a S#1:no$H_2$ Cu foil showing that Cu is oxidized in depth.

## 8) Energy-dispersive X-ray spectrometry spectra

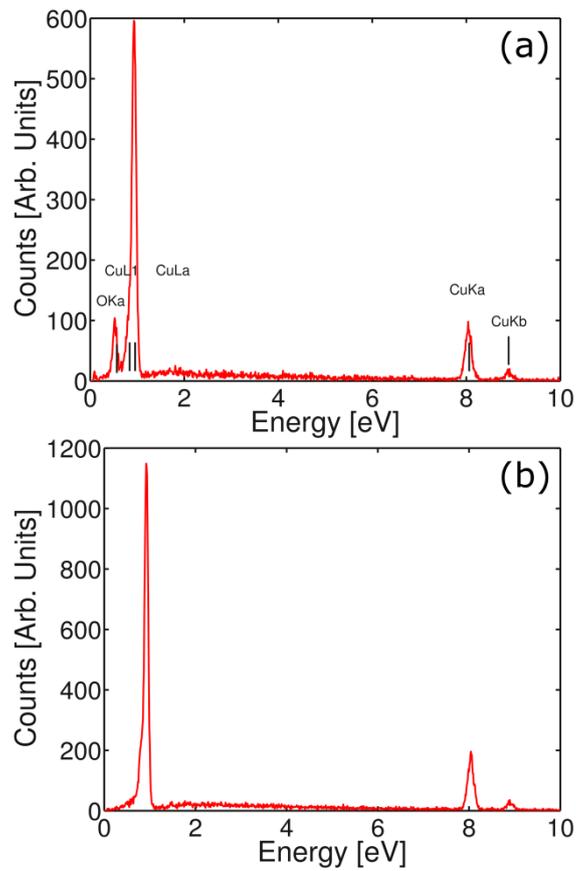

**Figure S7**: Energy-dispersive X-ray spectrometry spectra recorded on a Cu specimen subjected to S#1:noH$_2$. Spectrum of (a) a Cu$_2$O(111) inclusion and (b) the background.

## 9) Estimation of the oxygen partial pressure

First, we ensure that the flow inside the quartz tube is laminar. In that case, when argon is introduced, all the air in the tube is driven away without turbulence, gently pushed outside by argon. Therefore, argon does not mix with air, preserving its initial purity. The Reynolds number, Re, is a dimensionless number allowing to determine, given the rate and the geometry of the flow, if the regime is laminar or turbulent. For a flow in a pipe, it is given by the well-known formula:

$$Re = (\rho \times v_{moy} \times L)/\mu,$$

where $\rho$ is the density of the fluid [kg/m$^3$], $v_{moy}$ is the average velocity of the flow [m/s], L is a characteristic linear dimension [m], and $\mu$ is the dynamic viscosity of the fluid. In the case of a cylindrical pipe, the characteristic linear dimension is the diameter $\phi$, L = $\phi$.

The average velocity can be easily calculated: $v_{moy}$ = F/S, with F the volumetric flow rate [m$^3$/s] and S the section of the pipe [m$^2$]. For a cylindrical pipe, S = $\pi \times (\phi/2)^2$ and Re = $(4/\pi) \times (\rho \times F)/(\mu \times \phi)$.

In the specific case of an argon flow, $\rho$ = 1.7832 kg/m$^3$ and $\mu$ = 2.2×10$^{-4}$ kg/(m.s) (at 15 °C and 1013 mbar). We take into consideration the "worst case scenario" by taking the highest F value of 2 l/min (= 1/30×10$^{-3}$ m$^3$/s) used in the process. Finally, with $\phi$ = 2.5×10$^{-2}$ m, we find Re = 13.8. This value is well below the value of 2300 generally considered for a transition to a turbulent flow.[10] Here again, we consider an upper bound for Re with respect to the temperature, since we consider the room temperature values of $\rho$ and $\mu$, and $\rho$ varies as ~1/T while $\mu$ increases with increasing temperatures (~T$^{1/2}$), thus Re(1050 °C) << Re(15 °C).

In the case of a mixture of argon and hydrogen, the contribution of hydrogen can be neglected because it has a much smaller density compared with argon. In addition, the typical hydrogen flow used here (20 ml/min) is also much lower compared to the argon flow.

Based on the data provided by the argon cylinder supplier (Air liquide, purity alpha2 99.9995%), the amount of O$_2$ should be below 0.5 ppm. So we can estimate that the O$_2$ partial pressure $p_{O_2}$ is not greater than ~10$^{-7}$ bar. Based on the Ellingham diagram for the Cu/O$_2$ couple and considering $p_{O_2}$ = 10$^{-7}$ bar, we deduce that Cu$_2$O is stable for any temperature lower than ~1050 °C. This is all the more true for any $p_{O_2}$ greater 10$^{-7}$ bar.

## 10) Cu foil annealing in argon alone

Cu foils are annealed in argon alone for 30 min, 1 h or 2 h to observe how the morphology of the sample evolves with the hydrogen-free annealing time. From Figure 2e-g of the main manuscript, it is seen that, obviously, the density in $Cu_2O(111)$ inclusions increases with the exposure duration to residual oxygen. These crystals are preferentially formed on the rolling striations, where Cu shows more defects, or along grain boundaries. More importantly, when comparing the surface morphology of each sample, it appears clearly that the average grain size stagnates at ~50 µm and does not increase after 30 min, even if a prolonged annealing of 2 h is performed, as though the recrystallization becomes somehow inhibited. In a second experiment, the same Cu piece is annealed twice in a row in the same conditions (S#1:noH$_2$). The inspection of the foil at the exact same place by scanning electron microscopy discloses that, as could be anticipated based on the previous observation, the grain morphology does not evolve in the least bit (see Figure S8a,b). The only notable difference lies in the larger number of inclusions adding to the ones already present.

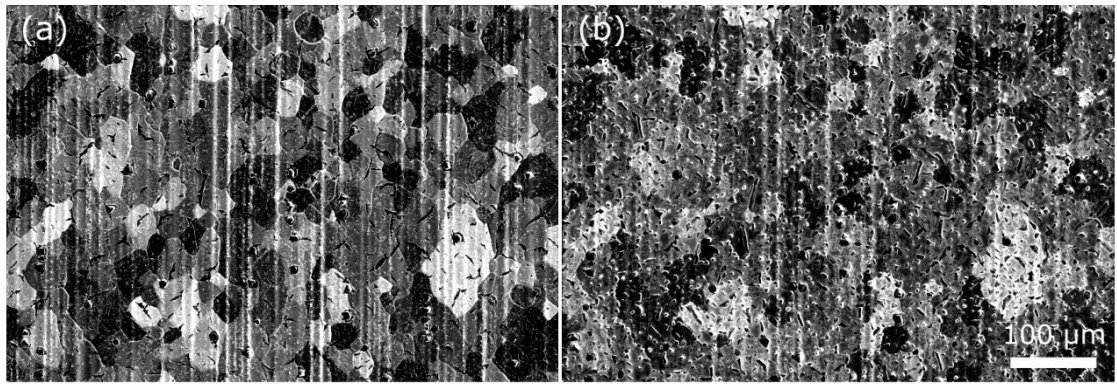

**Figure S8**: (a) View by scanning electron microscopy of the surface of a Cu foil annealed 30 min in argon alone and (b) the same foil at the same spot annealed a second time 30 min in argon alone.

## 11) Reverse pregrowth treatment

We evaluate the effect of a reverse pregrowth treatment (S#1:$H_2$/S#2:no$H_2$) on the growth of graphene. For the same methane flow as the one used in Figure 3c of the main text (0.35 sccm), the growth following that treatment exhibits a full coverage by monolayer graphene with many multilayer graphene patches underneath (see Figure S9a). Figure S9b displays a scanning electron microscopy picture for an amount of methane decreased down to 0.2 sccm in order to obtain a partial coverage (in the same methane flow conditions, no growth occurs with the standard S#1:no$H_2$/S#2:$H_2$ treatment). Even when the methane flow is decreased in that way, the coverage is almost complete, except for a few areas where it is possible to distinguish the edges of some hexagons only partially incorporated into the graphene film, or very occasionally isolated ones. The reduced methane flow allows determining that the typical hexagon size is below 50 μm. It is also worth noting that the Cu film is polycrystalline and the graphene flakes grow with random orientations.

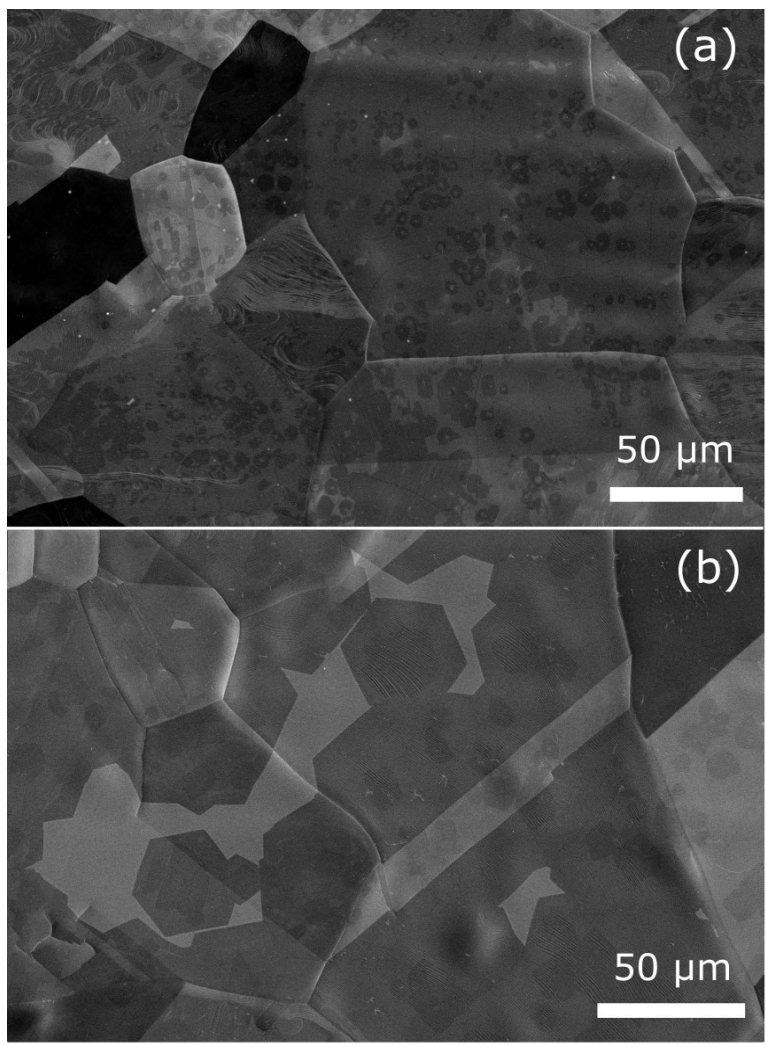

**Figure S9**: Scanning electron microscopy images of the graphene growth subsequent to the reverse pregrowth sequence, with a 0.35 (a) or 0.2 (b) methane flow.

## 12) Smoothing of the Cu foil after hydrogen annealing

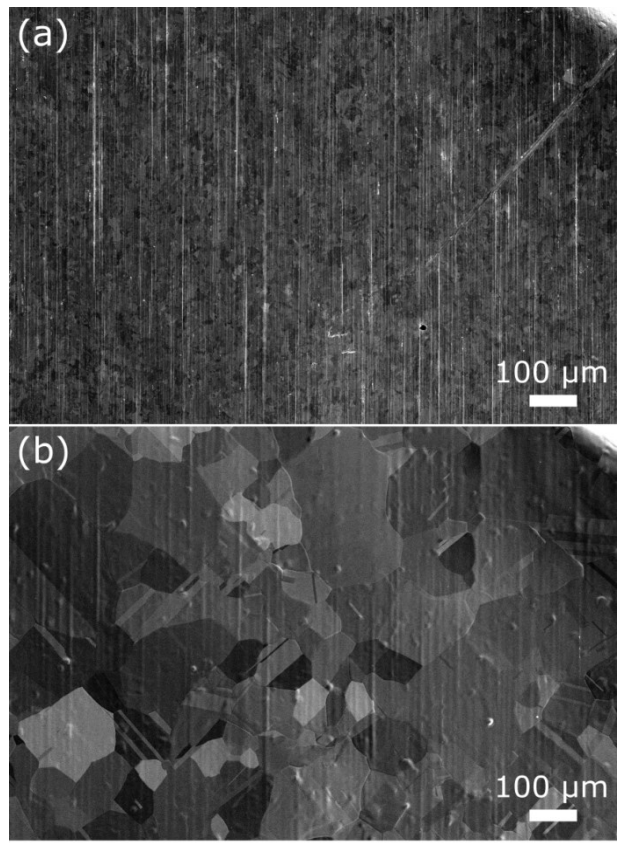

**Figure S10**: Scanning electron microscopy pictures of a Cu piece (a) before any thermal annealing and (b) subjected to the S#1:H$_2$/S#2:H$_2$ treatment at the exact same location. One can visualize the smoothing of the cold-roll striations and of a scratch in the top right corner of the foil. In addition, the annealed sample reveals a drastic increase of the grain size.

## 13) Supplementary low-energy electron diffraction patterns

The graphene/Cu(111) specimen seen in Figure 3c of the main text is analyzed in several points randomly chosen all over its surface (~1 cm$^2$). It is found out that the sample presents a unique Cu(111) orientation. Graphene shows epitaxial alignment with respect to Cu(111), as testified by the unique set of LEED spots in Figure S11a–c. A weak misorientation is occasionally observed, as attested by the very faint ring in Figure S11d (white arrow), possibly related to the fact that the foil is not polished, thereby causing little graphene misorientation with respect to Cu(111).[7]

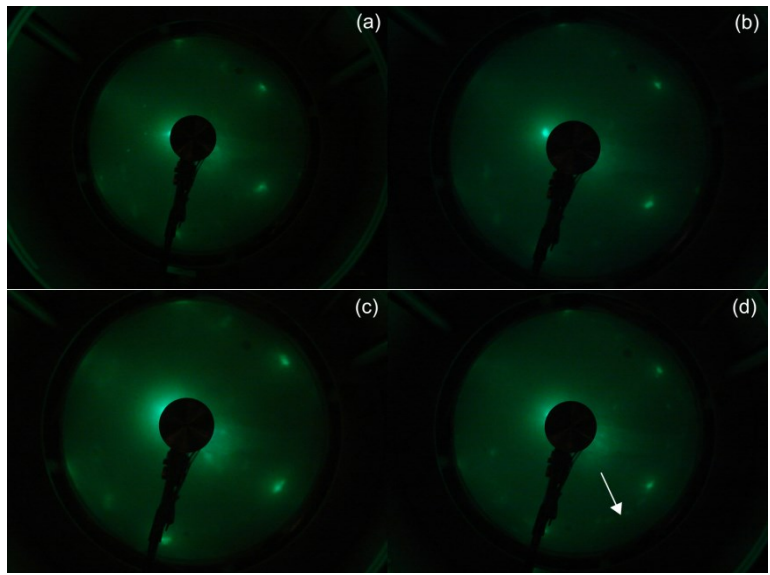

**Figure S11**: Low-energy electron diffraction patterns recorded in random positions on the sample pictured in Figure 3c of the main text.

## 14) Micro-Raman spectroscopy statistics

In the average and standard deviation values given in the main text, we also include the data recorded on the edges of the hexagonal graphene domain. These data often show a large spread relative to the average value. Indeed, by excluding the edges, the standard deviation decreases. The average and standard deviation are calculated from more than 400 data points.

|  | 2D-band FWHM [cm$^{-1}$] | 2D-band shift [cm$^{-1}$] | G-band shift [cm$^{-1}$] | $I_{2D}/I_G$ |
|---|---|---|---|---|
| µ | 25.8 | 2684.6 | 1582 | 2.3 |
| µ(wo edges) | 25.5 | 2684.4 | 1.1 | 0.5 |
| σ | 1.4 | 0.9 | 1585.3 | 2.3 |
| σ(wo edges) | 1.1 | 0.8 | 1 | 0.2 |

**Table S2**: Summary of the average and standard deviation of the 2D-band full width at half maximum, 2D-band shift, G-band shift, and 2D-band over G-band intensity ratio ($I_{2D}/I_G$) extracted from the micro-Raman mappings in Figure 4 of the main manuscript.

## 15) Nanobeam electron diffraction patterns

Figure S12 shows all the electron diffraction (ED) patterns recorded in nanobeam mode on the graphene flake shown in Figure 5a of the main text. All the ED patterns correspond to a graphene single crystal oriented along the [0001] zone axis. Between all the ED patterns, the rotation of the δ angle (defined in Figure 5b of the main text) is small, thus highlighting the single crystalline nature of the graphene flake.

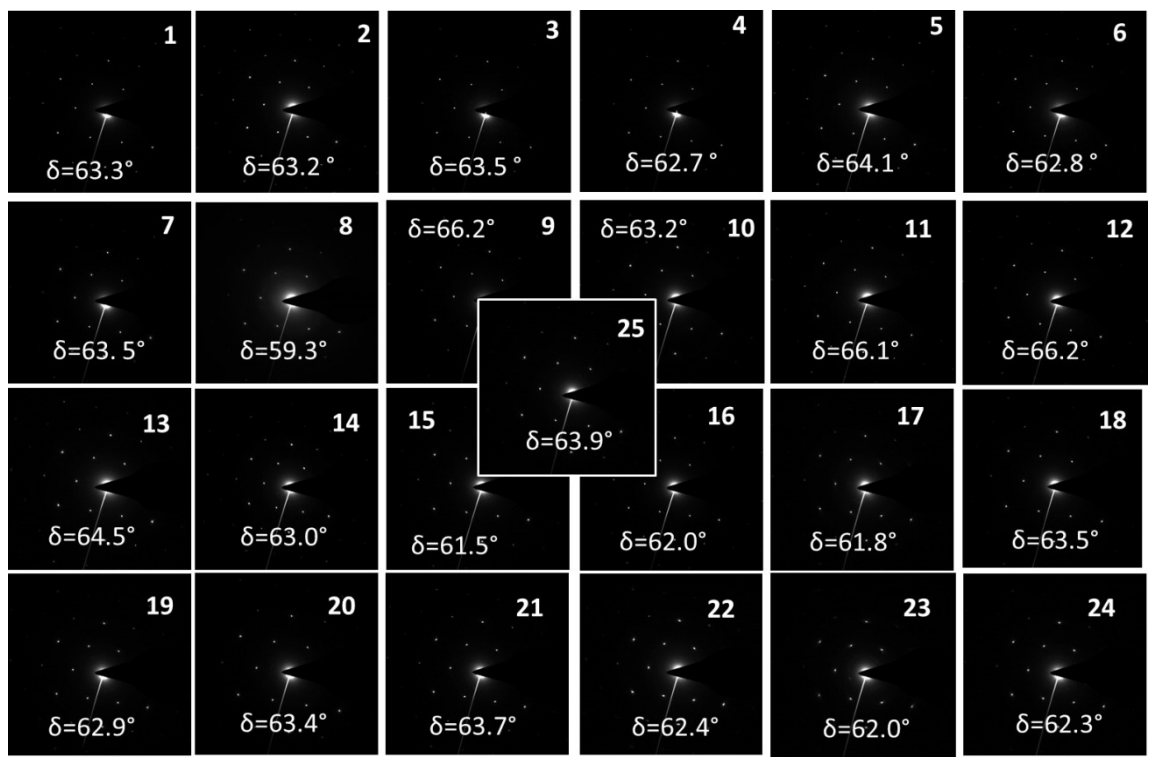

**Figure S12**: The 25 electron diffraction (ED) patterns recorded in nanobeam mode and through the holes of the TEM grid. Each ED pattern is labelled according to the location indicated in Figure 5a of the main text. In addition, the value of the δ angle (defined in Figure 5b of the main text) is given for each ED pattern.

## 16) Cleaning of the Cu foils

It is of great importance to properly prepare the Cu foils before growing graphene on top of it in order to achieve graphene films of the best quality. In the scientific literature, many different pretreatment techniques can be found: (1) chemical treatment in various liquids such as acetic acid,[11,12,13,14,15,16,17] solvents,[18] water,[19,20] inorganic acids (dilute $HNO_3$,[11,19,21] dilute HCl,[21,22,23]), $FeCl_3$,[11,24] Cr or Ni etchants;[21] (2) electropolishing;[11,12,13,19,25] (3) chemical mechanical polishing.[26] More particularly, the removal of Cu oxide with acetic acid was shown to be very effective.[27] In the perspective of large-scale production, it also presents the distinctive asset not to involve complicated treatments or hazardous chemicals. On the other hand, many works reported in the literature deal with graphene growth on Alfa Aesar Cu foils (25-μm-thick; 99.8% purity; reference number 13382; explicitly mentioned in[11,12,14,15,18,19,20,23,24,25,28,29]). However, it is rarely reported that these foils are in fact coated with a thin metallic oxide anticorrosion film.[16,25,29] It goes without saying that this coating must be entirely removed, without degrading the underlying Cu foil, prior to graphene synthesis.

Here, we consider the widely used Cu foils mentioned above (Alfa Aesar #13382). X-ray photoelectron spectroscopy was first used to assess the presence of contaminants on the foil's surface before any treatment (so-called "as-received" foil). Two different as-received samples (1×1 $cm^2$) are analyzed in two and three distinct spots randomly chosen, respectively. The concentration of the detected elements is determined from the survey scan (see Figure S13). The main detected elements are oxygen (O 1$s$) and carbon (C 1$s$), corresponding to organic contamination. The presence of chromium (Cr 2$p$) is also unambiguously identified, corresponding to a layer of chromium oxide (the anticorrosion coating mentioned above). In addition, peaks that can be attributed to calcium (Ca 2$p$; between 8 and 10%) and phosphorus (P 2$p$; around 6%) appear. Other contaminants such as nitrogen (N 1$s$) or chlorine (Cl 2$p$) are also found, albeit in very weak amounts (less than 1.5%). The concentrations in the different elements can be found in Table S3.

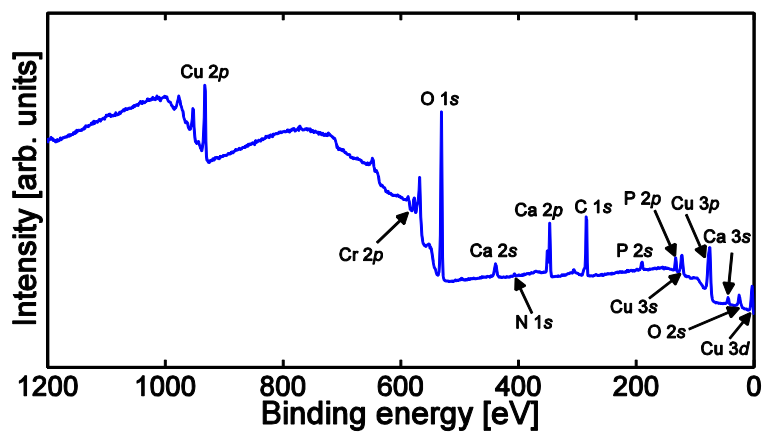

**Figure S13**: X-ray photoelectron spectroscopy survey scan of an as-received Cu sample.

In order to remove the observed contaminants, we have next tested simple recipes involving harmless chemicals such as glacial acetic acid (GAA; Acros Organics; >99.8% purity), distilled water (DW) or isopropyl alcohol (IPA): (1) GAA alone for 15 min; (2) DW alone for 15 min; (3) mixture of GAA and DW in a 1:1 ratio for 15 min; (4) GAA alone for 15 min followed by rinsing in IPA. All the treatments are performed at room temperature. Finally, the Cu pieces are gently blown dry with nitrogen. All the recipes leave the chromium oxide layer intact, as one would expect (see Figure S14a). The only technique being efficient in removing calcium is the third one (GAA+DW) (see Figure S14b). Phosphorus is still present in all cases. The concentrations in the different elements are summarized in Table S3.

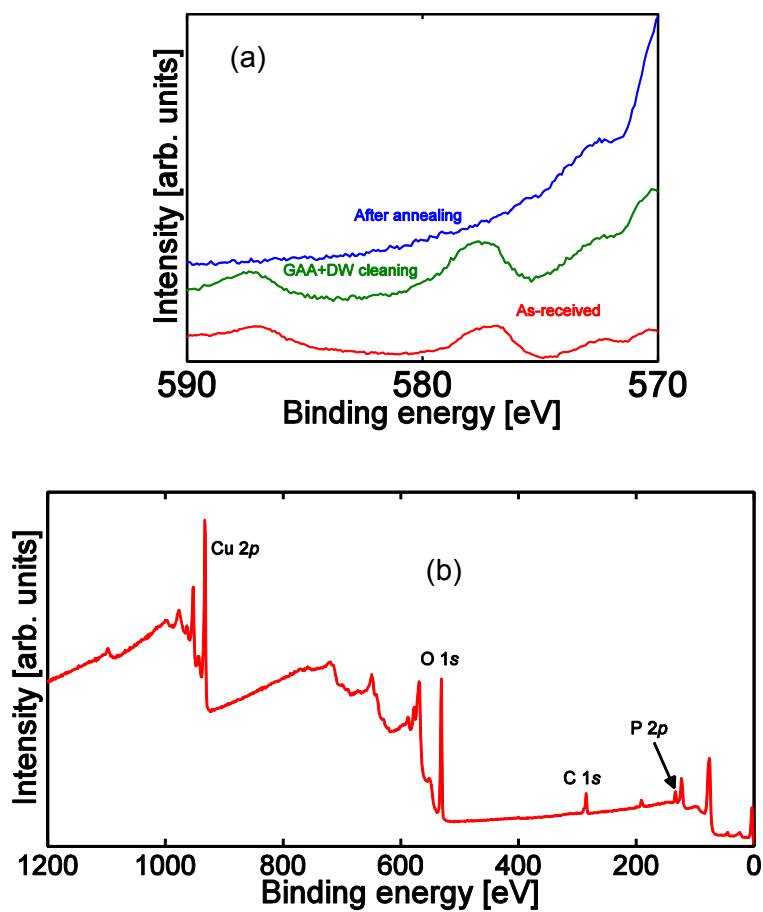

**Figure S14**: (a) Cr 2*p* core level spectrum for the as-received, the GAA+DW cleaned, and the annealed Cu pieces, respectively. (b) X-ray photoelectron spectroscopy survey scan of a Cu sample cleaned in a mixture of glacial acetic acid and distilled water.

|  |  | Composition | | | | | | | |
| --- | --- | --- | --- | --- | --- | --- | --- | --- | --- |
| Sample type | Point | O 1$s$ | Cu 2$p$ | C 1$s$ | Ca 2$p$ | P 2$p$ | Cr 2$p$ | N 1$s$ | Cl 2$p$ |
| As- received #1 | 1 | 40 | 5 | 38.5 | 8 | 5.6 | 1.4 | 1.5 | X |
|  | 2 | 41.8 | 5 | 35.9 | 8 | 6.1 | 1.5 | 1.7 | X |
| As- received #2 | 1 | 46.9 | 5.9 | 28.9 | 9.4 | 6.4 | 2 | X | 0.5 |
|  | 2 | 46.7 | 5.8 | 29 | 9.7 | 6.4 | 1.8 | X | 0.6 |
|  | 3 | 45.3 | 5.8 | 30.3 | 9.5 | 6.5 | 1.8 | X | 0.8 |
| Cleaned in GAA | 1 | 51.1 | 13.7 | 16.5 | 6.6 | 9 | 3.1 | X | X |
|  | 2 | 51.8 | 13.2 | 16.3 | 6.4 | 9 | 3.3 | X | X |
|  | 3 | 51.3 | 14.2 | 14.1 | 7.2 | 9.8 | 3.4 | X | X |
| Cleaned in DW | 1 | 47 | 10.8 | 24.8 | 6 | 8.8 | 2.6 | X | X |
|  | 2 | 46.2 | 9.4 | 28.6 | 5.2 | 7.9 | 2.7 | X | X |
|  | 3 | 45.2 | 9.5 | 28.8 | 5.5 | 8.4 | 2.6 | X | X |
| Cleaned in GAA+DW | 1 | 48.7 | 17.6 | 22.3 | X | 8.2 | 2.6 | X | 0.6 |
|  | 2 | 49.2 | 17.9 | 21.4 | X | 8.1 | 2.8 | X | 0.6 |
|  | 3 | 49.6 | 22.2 | 14.8 | X | 9.8 | 2.9 | X | 0.7 |
| Cleaned in GAA, rinsed in IPA | 1 | 52.1 | 12.6 | 17.6 | 6.3 | 8.5 | 2.9 | X | X |
|  | 2 | 52.3 | 12.1 | 17.2 | 6.8 | 8.5 | 3 | X | X |
|  | 3 | 50.9 | 12.2 | 18.2 | 6.7 | 9.1 | 2.9 | X | X |

**Table S3**: Superficial composition of the as-received samples and of the samples cleaned by the four different methods, obtained from the survey scan.

We have next investigated the superficial composition after annealing (i.e. growth without methane, in the standard conditions, see Figure 1a of the main text) or after growth (in the standard conditions as well), for each type of treatment, in three distinct points. Table S4 summarizes the composition of the surface of the diverse Cu pieces. The conclusions are essentially the same in both cases: the chromium oxide layer evaporates during the thermal treatment (also confirmed by the high-resolution core level spectrum of Cr 2$p$ in Figure S14a), as well as phosphorus, chlorine, and nitrogen, for all four treatments. The only remaining impurity is calcium, in small amounts (around 1%), except for the GAA+DW recipe, as already observed before. In addition, the only treatment leading to the successful growth of graphene is GAA+DW, as testified by the much higher carbon concentration of the sample cleaned with GAA+DW (~48% versus less than 15% for the other three methods) and careful scanning electron microscopy (SEM) inspection of each kind of samples (not shown). This is further illustrated by the SEM picture displayed in Figure S15, corresponding to a growth performed on a sample half-dipped in the GAA+DW mixture. It clearly shows that graphene grows on the half dipped in GAA+DW while it does not on the untreated one. It is even possible to perceive a difference in color between both parts, evidencing the effect of the GAA+DW cleaning.

In conclusion, it appears (1) that the inhibited growth of graphene is related to the imperfect removal of calcium from the surface and (2) that the chromium oxide layer is removed by the thermal treatment (and in fact even before the beginning of the growth process, since otherwise, the growth would not occur), as well as the other minor contaminants. This study also illustrates that some cleaning recipes presented in the literature are not universal and the cleaning of the Cu foil must be adapted to the foil's manufacturer (coating, contaminations).

| Sample type | Point | Composition | | | | | | | |
|---|---|---|---|---|---|---|---|---|---|
| | | O 1s | Cu 2p | C 1s | Ca 2p | P 2p | Cr 2p | N 1s | Cl 2p |
| Cleaned in GAA/after annealing | 1 | 37.2 | 46.3 | 15.1 | 1.4 | X | X | X | X |
| | 2 | 35 | 51.8 | 11.8 | 1.4 | X | X | X | X |
| | 3 | 36.2 | 48.2 | 14.5 | 1.1 | X | X | X | X |
| Cleaned in DW/after annealing | 1 | 37.5 | 45.6 | 16.1 | 0.8 | X | X | X | X |
| | 2 | 34.7 | 51.8 | 12.8 | 0.7 | X | X | X | X |
| | 3 | 36.8 | 48.1 | 14 | 1.1 | X | X | X | X |
| Cleaned in GAA+DW/after annealing | 1 | 37.7 | 46.1 | 16.2 | X | X | X | X | X |
| | 2 | 35.8 | 52.5 | 11.7 | X | X | X | X | X |
| | 3 | 36.5 | 50.9 | 12.6 | X | X | X | X | X |
| Cleaned in GAA, rinsed in IPA/after annealing | 1 | 37.4 | 44.5 | 17.1 | 1 | X | X | X | X |
| | 2 | 37.2 | 47.8 | 13.9 | 1.1 | X | X | X | X |
| | 3 | 35.9 | 48.2 | 14.7 | 1.2 | X | X | X | X |
| Cleaned in GAA/after growth | 1 | 36.7 | 45.2 | 15.8 | 2.3 | X | X | X | X |
| | 2 | 29.8 | 61.2 | 7 | 2 | X | X | X | X |
| | 3 | 25.7 | 68.3 | 3.7 | 2.3 | X | X | X | X |
| Cleaned in DW/after growth | 1 | 35 | 50.6 | 13.7 | 0.7 | X | X | X | X |
| | 2 | 27.5 | 65.5 | 6.6 | 0.4 | X | X | X | X |
| | 3 | 29.1 | 61 | 9.4 | 0.5 | X | X | X | X |
| Cleaned in GAA+DW/after growth | 1 | 5.7 | 46.2 | **48.1** | X | X | X | X | X |
| | 2 | 6 | 45.9 | **48.1** | X | X | X | X | X |
| | 3 | 6 | 45.7 | **48.3** | X | X | X | X | X |
| Cleaned in GAA, rinsed in IPA/after growth | 1 | 38.8 | 43.2 | 16 | 2 | X | X | X | X |
| | 2 | 37.8 | 48.6 | 7 | 1.9 | X | X | X | X |
| | 3 | 34.6 | 54.1 | 8.7 | 2.6 | X | X | X | X |

**Table S4**: Superficial composition of the samples cleaned by the four different methods after annealing or after growth, obtained from the survey scan.

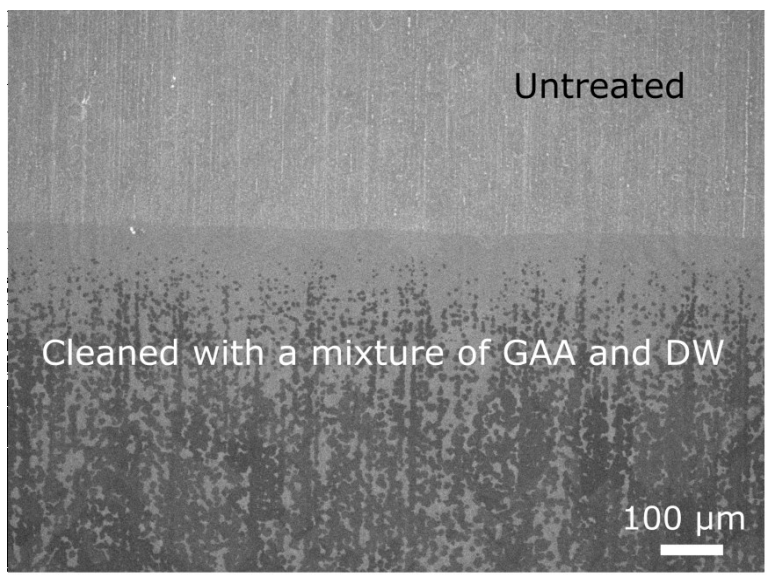

**Figure S15**: Scanning electron microscopy picture corresponding to a growth performed on a Cu piece half-dipped in the GAA+DW mixture.

# 17) Additional experimental details

**Scanning electron microscopy (SEM)/Energy-dispersive X-ray spectrometry (EDX)**:

The morphology and the size of the Cu grains and graphene domains are monitored with two different microscopes: a Jeol JSM-6010LV InTouchScope at low magnification (operated at an accelerating voltage of 5 kV and a spot size between 30 and 50, with a working distance of 25 mm to increase the field of view, in secondary electron mode) and a Jeol JSM-7500F at high resolution (operated at an accelerating voltage of 1 kV and an emission current of 5 μA, with a working distance of 3 mm, in secondary electron mode with a low gentle beam of 0.2 kV applied to the specimen). EDX mapping is performed with the Jeol JSM-7500F at 15 kV with a probe current of 1 nA and a resolution of 512×384 px.

**Electron-backscattering diffraction (EBSD)**:

The analysis of the Cu foil crystal orientation is performed using a SEM ZeissSupra55 fitted with a HKL-Oxford Instruments EBSD system featuring the Nordlys II camera. Data analysis is realized using the associated Channel 5 software suit. EBSD data collection is operated at an accelerating voltage of 15 kV, a working distance of 11 mm, and a sample tilt of 70°. The out-of-plane inverse pole figure (IPF) maps are in the $z$ direction, perpendicular to the Cu foil surface, while the in-plane IPF maps are in the $y$ direction, parallel to the plane of the Cu foil (instead, the in-plane IPF maps could also be given in the $x$ direction, the information being the same). The zero solutions (electron-backscattered diffraction patterns that cannot be indexed due to very poor quality) are replaced by an extrapolation based on the neighboring points.

| Figure | Pixel size [μm$^2$] | Raster size | % of correct indexation |
|---|---|---|---|
| 2a | 4×4 | 100×100 | 84 |
| 2c,f | 10×10 | 194×148 | 97.45 |
| 2d,g | 13×13 | 100×100 | 95.64 |
| 2e,h | 10×10 | 197×143 | 86.86 |
| S4 | 2×2 | 150×114 | 96.78 |
| S5 | 12×12 | 261×194 | 95.68 |

**Table S5**: More details about the different inverse pole figure maps presented in this work.

**Low-energy electron diffraction (LEED)**:

The LEED patterns are acquired in a ultrahigh-vacuum setup after outgassing the samples at 300 °C for one hour. The energy of the incident electrons is set to 70 eV, with an analysis spotsize of ~1 mm.

**X-ray photoelectron spectroscopy (XPS)**:

A ThermoFisher Scientific K-alpha spectrometer is utilized. It is fitted with a monochromatized Al $K\alpha$1,2 x-ray source and a hemispherical deflector analyzer. The spectra are recorded at constant pass energy (150 eV for depth profiling and survey; 30 eV for high resolution spectra). A flood gun (low energy electrons and Ar$^+$ ions) is used during all the measurements. During the sputtering, the Ar$^+$ ion gun is operated at an accelerating voltage of 2 kV, with an erosion time of 5 s per cycle, and the analysis is done in snapshot mode. The XPS data are treated with the Avantage software. High resolution spectra are fitted by Gaussian-Lorentzian lineshapes with an Avantage "smart" background (*i.e.* a Shirley background in most cases, or a linear background in case the lineshape decreases with increasing binding energy). The diameter of the analyzed surface is 250 μm.

**Micro-Raman spectroscopy (μRS)**:

A LabRam HR 800 confocal laser system from Horiba Jobin Yvon is used for the acquisition of the Raman spectra. The measurements are performed at room temperature with a laser wavelength of 514 nm in backscattering geometry. The spectra are acquired on a 950 μm by 1100 μm grid with a spacing of 30 μm between each measurement spot (totalizing 32×37 spectra). A 100× objective (NA = 0.95) is used to collect the signal. The incident power is kept below 1 mW to avoid any heating effect. High resolution (1800 lines/mm) gratings are used for the measurements (with a corresponding spectral resolution < 1 cm$^{-1}$). For the data analysis, the spectra are first fitted with Lorentzian functions. Then, the full width at half maximum, the position, and the integrated intensities of the 2D and G bands are extracted from the resulting fits.

**Transmission electron microscopy (TEM)/electron diffraction (ED)**:

The TEM experiments are performed by using a FEI Titan Cube operating at 80 kV and equipped with a $C_S$ image corrector. The analyses are performed at liquid-nitrogen temperature to hinder carbon contamination and electron-beam damages. The effective diameter of the area probed on the specimen during nanobeam electron diffraction is around 150 nm.